# A Spin-Optical Nano Device

Enno Krauss, Gary Razinskas, Dominik Köck, Swen Grossmann, and Bert Hecht*

*Nano-Optics and Biophotonics Group, Department of Experimental Physics 5, Wilhelm-Conrad-Röntgen-Center for Complex Material Systems (RCCM), University of Würzburg, Am Hubland, 97074 Würzburg, Germany*

*Email: hecht@physik.uni-wuerzburg.de

**The photon spin is an important resource for quantum information processing as is the electron spin in spintronics. However, for subwavelength confined optical excitations, polarization as a global property of a mode cannot be defined. Here, we show that any polarization state of a plane-wave photon can reversibly be mapped to a pseudo-spin embodied by the two fundamental modes of a subwavelength plasmonic two-wire transmission line. We design a device in which this pseudo-spin evolves in a well-defined fashion throughout the device reminiscent of the evolution of photon polarization in a birefringent medium and the behaviour of electron spins in the channel of a spin field-effect transistor. The significance of this pseudo-spin is enriched by the fact that it is subject to spin-orbit locking. Combined with optically active materials to exert external control over the pseudo-spin precession, our findings could enable spin-optical transistors, i.e. the routing and processing of quantum information with light on a subwavelength scale.**

If photons are confined - in particular to subwavelength scales - significant longitudinal fields do emerge in order to satisfy Maxwell's equations. As a result, the direction of the electric field vector varies significantly over the mode profile. This complicates the notion of polarization as orientation of the transverse electric or magnetic field vector of a mode. Nevertheless, nanostructures can exhibit a strongly polarization-dependent scattering of light into plane-wave or guided modes. Recently, this effect, called spin-orbit locking, attracted considerable attention[1–10]. It describes the pronounced spin-dependent directional scattering from nanostructures illuminated by circularly polarized plane waves[1–6] or by emitters exhibiting a circularly polarized dipole moment which couples to the evanescent field of guided modes[7–10]. In both cases the spin-orbit locking effect happens locally and instantaneously and then results in spin dependent propagation directions of the scattered modes. The scattered modes themselves no longer carry a defined spin. However, in view of possible applications, such as routing and processing of quantum information[11–13] at a subwavelength scale and subwavelength chiral sensing applications[14], it is highly desirable to realize optical nanocircuitry in which the photon spin remains a well-defined observable throughout the nanocircuit and even remains well-defined upon conversion into a far-field photon.

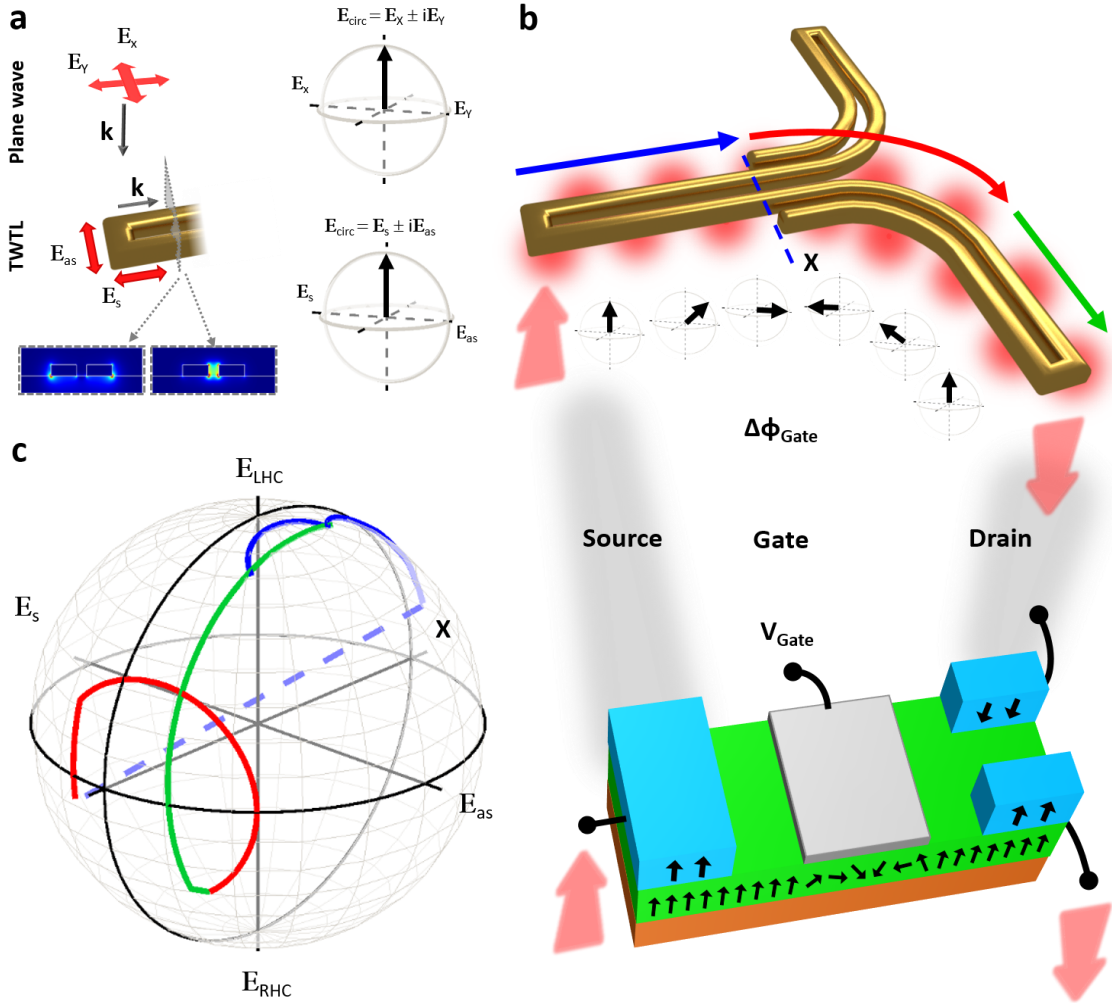

**Figure 1 | Spatial separation of photons according to their spin.** a) The polarization state of a photon can be mapped onto the two fundamental modes of a plasmonic waveguide. Bottom: Cross-sectional intensity distributions of the symmetric (left) and antisymmetric (right) TWTL eigenmodes. Right: Spherical representation of the Stokes vectors, the Poincaré sphere, based on transverse electric field components (top sphere) or fundamental TWTL modes (bottom sphere). b) Sketch of the working principle of a spin-optical nano device, analogous to an electron spin-transistor, with in-coupling (source) and out-coupling (drain) regions as well as a gate-region, where the precision of the spin-state is determined by the accumulated phase (gate voltage). c) The trajectory of the photons pseudo-spin state throughout the whole device represented on a Poincaré sphere.

Here we demonstrate that it is possible to define a quantity analogous to the photon spin also for deeply subwavelength-confined optical modes, such as plasmonic excitations in a two-wire transmission line (TWTL). This is achieved by reversibly mapping the polarization state of a far-field photon – represented by a complex transverse electric field vector - to a superposition of two fundamental modes of a plasmonic TWTL[15] which we call a pseudo-spin (see Fig. 1a). We find that upon propagation the pseudo-spin vector is always defined and undergoes characteristic trajectories on a Poincaré sphere reminiscent of light propagation in a birefringent medium and the precession of electron spins in the channel of a spin field-effect transistor[16,17] - the hallmark of spintronics[18–20]. The idea of representing the evolution of a superposition of two coupled modes by a unit vector and a corresponding trajectory on a Poincaré sphere is already known[21]. However, in the present case, as opposed to the case of coupled-mode theory in photonics[22], it is well justified to designate the mode superposition vector as pseudo-spin since (i) conversion into a far-field photon of a corresponding real polarization state is possible at any time and (ii) because of the fact that spin-orbit locking can be imposed, as is discussed below. The notion of a pseudo-spin vector that develops in a well-defined way throughout a device in which pseudo-spin-dependent scattering may occur, justifies drawing an analogy to electron spin-based devices, such as a spin transistor (Fig. 1b).

To illustrate possible applications, we present a prototypical spin-optical plasmonic nano device with one input and two output ports all interfacing to propagating photons. The device sorts photons according to their spin and re-emits them in their initial spin-state. Utilizing the spin degree of freedom in nano-optical circuitry is expected to open new possibilities for subwavelength-integrated active spin-optical devices for local sensing applications as well as for the processing of quantum information at the nanometre scale.

A plasmonic TWTL, featuring two fundamental guided eigenmodes,[15] is the basic building block of the spin-optical nano device. Overall, the device consists of a Y-Junction of TWTLs as sketched in Figure 1b. At the input and output ports of the device antennas are used to convert polarized far-field photons into a superposition of TWTL modes and vice versa. The antennas are designed such that both amplitude and phase of the two transverse electric field vector components are converted into a superposition of waveguide modes with corresponding amplitude ratio and initial phase offset (see Fig. 1a, the methods section and supplementary information). This is achieved by a split-ring antenna design that provides equally improved coupling efficiencies for both modes compared to dipolar antennas[23]. A photon of a certain spin is therefore converted into a corresponding superposition of TWTL eigenmodes described by a pseudo-spin vector. During propagation of the mode superposition its pseudo-spin vector describes a trajectory on a Poincaré sphere (coloured paths in Fig. 1c). Propagation along straight TWTL segments (blue and green parts of the trajectory in Fig. 1b,c) is accompanied by the accumulation of a characteristic phase shift between the modes due to their different effective wavelengths, an effect similar to birefringence of a plane-wave photon in an anisotropic crystal. This causes a precession of the pseudo-spin vector reminiscent to that of electron spins in the gate region of an electron spin field-effect transistor. While for an electron spin transistor the amount of precession can be tuned by a gate voltage, in the TWTL it is determined by the accumulated phase shift of the two modes which is tuned passively via the waveguide length and cross section. While not yet implemented in this work, it is conceivable that active tuning might be achieved e.g. by placing optically active or nonlinear materials around the transmission line (see supplementary information). Figure 1c illustrates the corresponding initial part of the pseudo-spin trajectory (blue) on the Poincaré sphere starting with left circular polarization at the north pole down to a point X close to the equator corresponding to the end of the initial straight part of the TWTL (see also supplementary information). The path deviates slightly from a meridian due to the unequal Ohmic damping of the fundamental TWTL modes, similar to optical dichroism. Note that the starting point of the blue trajectory is slightly displaced from the north pole due to remaining small imbalances in the in-coupling intensity and phase of the antenna for each mode. Also note the small wiggles in the blue part of the polarization trajectory that are due to the formation of standing waves upon reflection of parts of the intensity carried by the modes at X (see supplementary information).

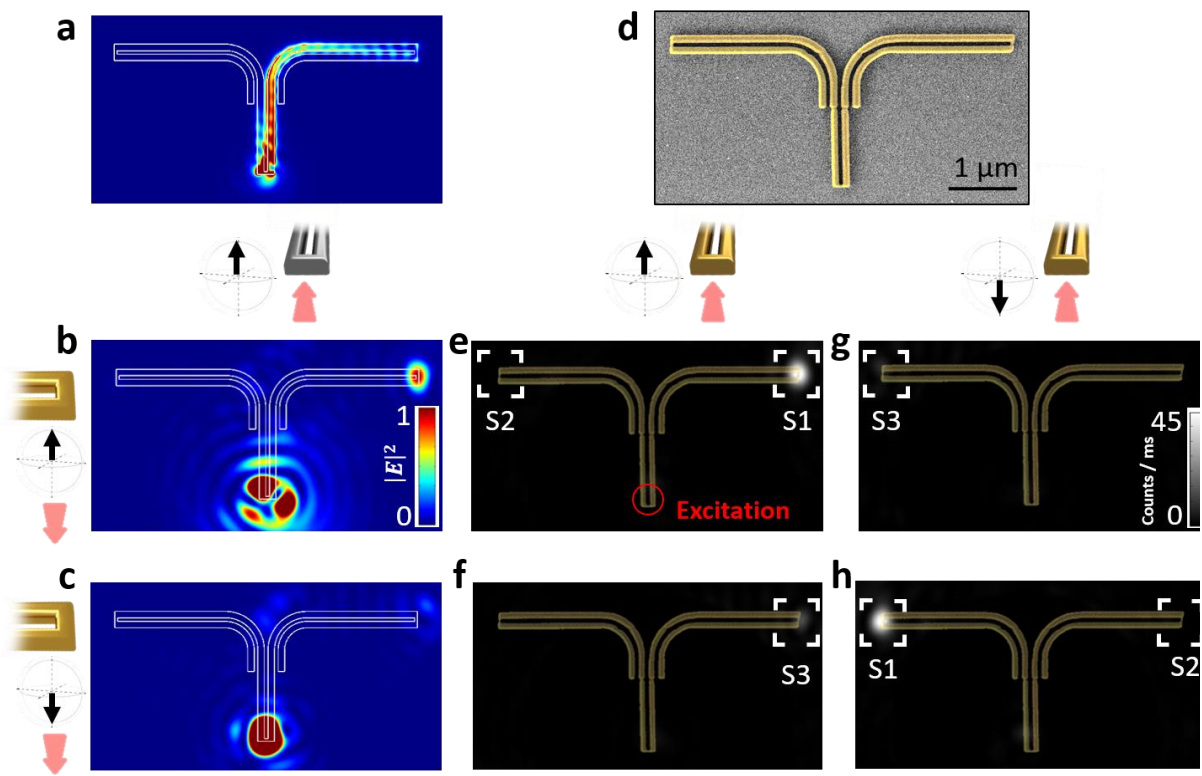

**Figure 2 | Numerical simulations and optical measurements.** Simulated near-field (a) and far-field (b,c) intensity maps for the optimized device geometry, where the left-handed pseudo-spin is excited. The far-field emission is analysed for left-handed (b) and right-handed (c) pseudo-spin contributions. d) Colourized scanning electron micrograph of the structure after transfer onto glass. CCD-images for excitation of the left-handed (e,f) or right-handed (g,h) pseudo-spin state with the analyser in the

detection path set to transmit only emission of the left-handed (e,g) or right-handed (f,h) pseudo-spin state, respectively. The structure's position is indicated by the overlaid SEM-image from d). The excited antenna is indicated in e).

X also denotes the transition from the initial straight TWTL portion to the Y-junction region (see Fig. 1b). In the Y-junction region two TWTLs are connected to the initially straight TWTL which, after a short segment of parallel propagation, are separated by a 90° bend of (see Fig. 1b). Due to the sharp bending the two emerging TWTL branches can be treated as uncoupled. In our device the initial TWTL length (blue trajectory in Fig. 1b,c) is chosen such that upon left-(right-)circularly polarized excitation spatial interference of the transverse mode profiles will lead to a field maximum only on the right (left) wire (see Fig. 2a and supplementary information). Therefore, upon circularly polarized excitation, only one of both paths at the Y-junction will carry intensity[1,24]. We thereby effectively invoke pseudo spin-orbit locking in the device. The following red part of the pseudo-spin trajectory in Figure 1c describes the pseudo-spin in the respective branch of the Y-junction. Due to lateral offset of the TWTL at X, a phase jump is introduced leading to a jump of the pseudo-spin onto the opposite side of the Poincaré sphere. Due to the bending of the TWTL, pronounced mode conversion occurs for the fundamental modes[25]. Some minor conversion to strongly damped, higher-order modes and to radiation does also occur (see supplementary information). The possibility to induce mode conversion is an important asset as it allows us to overcome the effect of the unequal Ohmic damping of the fundamental TWTL eigenmodes. Without mode conversion the pseudo-spin trajectory on the Poincaré sphere would continue to spiral away from the initial meridian and the poles of the sphere, corresponding to states of well-defined spin, could never be reached. Mode conversion in the curved part of the TWTL branch indeed causes the polarization trajectory to describe a closed loop. We note that this likely causes the accumulation of a Pancharatnam geometric phase[26,27] which does not influence the function of the present device, but may find applications in the future. As soon as the final straight part of the TWTL is reached (green trajectory in Fig. 1b,c) the pseudo-spin trajectory continues as discussed before, until it reaches the upper pole. At this point the respective eigenmode superposition is converted into a photon of well-defined spin by the out-coupling antenna.

**Results**

Full-3D finite-difference time-domain (FDTD) simulations were performed to optimize the geometry of the device for both effective sorting of photons according to spin as well as restoring the initial pseudo-spin state at the design wavelength of 800 nm. To accommodate geometrical deviations during sample fabrication, a corresponding 2D array of structures was fabricated by focused ion-beam milling of a single-crystalline gold platelet. For a fixed TWTL-cross-section and bending radius, the length of the linear TWTL segments (blue and green parts of the trajectory in Fig. 1b,c, respectively) have been varied (see supplementary information). The structure that showed the best performance was chosen for final simulations (Fig. 2a-c) and experiments (scanning electron microscopy (SEM) image in Fig. 2d). The measurements (Fig. 2e-h) were performed with a home-built inverted microscope, where the spin-dependent excitation and analysis were carried out with a quarter-wave plate and linear polarizers (see methods).

In FDTD simulations the structure's resulting far-field intensity for an excitation with left-handed circular polarized photons (from the viewpoint of the source) is analysed according to the spin of emitted photons. Considering only photons with left-handed circular polarization, corresponding to emission from the initially excited pseudo-spin state, indeed confirms emission from only the right output terminal (in top-view) of the structure (Fig. 2b). Filtering the same far-field simulation for the right-handed circular components instead (Fig. 2c), reveals only negligible intensities at either of the two output ports. This confirms the preservation of the emitted pseudo-spin state. In both images nearly no light is emitted from the left part of the structure, confirming the highly effective spin sorting.

In the experiment the structure was excited at the in-coupling antenna using a tightly focused (NA = 1.4) laser beam with a wavelength of 800 nm. Corresponding CCD-images for an excitation with left-handed and right-handed circularly polarized photons are presented in Figures 2e,f and g,h, respectively. For Figure 2e and h, emitted light with the spin matching the excitation is detected. The recorded image for an excitation with left-handed circularly polarized photons (Fig. 2e) reproduces the simulated far-field image in Figure 2b. The corresponding image for right-handed circularly polarized excitation (Fig. 2h) confirms the switching. The images in Figure 2f and g have been analysed for the spin opposite to the excitation, and thus emission of opposite spin photons. The negligible detected intensity confirms the absence of contributions of such photons at the output terminals of the device.

To quantify the sorting contrast we defined a figure of merit: $SC_{L,R} = I_{S1} / (I_{S1} + I_{S2})$ for the left-handed (L) and right-handed (R) circular excitation. The intensities $I_{S1}$ and $I_{S2}$ were extracted by integrating over a square of 500 nm around the chosen antenna (Fig. 2e,h) in the measurements filtered for the exciting polarization. For the left-handed (right-handed) circular excitation we achieve $SC_{L(R)} = 0.968 \pm 0.002$ ($0.979 \pm 0.001$), which is close to the simulated value $SC_{Sim} = 0.973$ and confirms nearly perfect spatial spin sorting. Here and in the following the errors were obtained by shifting the integration frame by two pixels. To characterise how well the spin is preserved, we keep the excitation fixed and record the output emission for the opposite spin. The CCD images for left- and right-handed circularly polarization excitation have been obtained with the analyser set to transmit the same (Fig. 2e,h) or opposing (Fig. 2f,g) spin, corresponding to the initial and opposing pseudo-spin state, respectively. We define the polarization contrast for the left-handed circular excitation $PC_L = I_{S1}/(I_{S1}+I_{S3})$ where S3 represents the white square in Figure 2f. $PC_R$ was extracted accordingly from the top left end of the structure from Figures 2g,h. We obtain $PC_{L(R)} = 0.815 \pm 0.005$ ($0.870 \pm 0.009$), close to the simulated value $PC_{Sim} = 0.868$.

In conclusion, we have demonstrated that any polarization state of a far-field photon can be translated to the subwavelength scale by mapping it to a pseudo-spin vector consisting of the amplitudes and phases of two plasmonic eigenmodes of a TWTL. To demonstrate the power of the approach we developed and realised a subwavelength plasmonic spin-optical nano device capable of spatially sorting photons according to their spin and re-emitting them with the same spin. Spin sorting is achieved by inducing spin-orbit locking through taking advantage of the spatial interference of the propagating modes.

Combinations of straight and curved TWTL elements enable us to fully control the resulting pseudo-spin trajectory on a Poincaré sphere in order to adjust the polarization of the emitted photon at will. Our findings open the road to the realisation of integrated spin-optical nano devices that will be of interest for building transistor-like elements for sensing as well as for the processing of quantum information on the nanoscale.

**Methods**

**Simulations:** All simulated results were obtained by FDTD simulations (Lumerical Solutions, FDTD Solutions). The nano device, situated on a glass-air interface ($n_{glass}$ = 1.51), is made from gold, whose wavelength dependent dielectric function is modeled by an analytical fit to experimental data[28,29]. The structure is excited from the substrate side by a tightly focused Gaussian source (λ=800 nm, numerical aperture NA=1.4). For a fixed antenna geometry, bending curvature and cross-section of the TWTL, the initial and final length of the structure were optimized to obtain the best spatial switching and polarization contrast. The mode distributions were obtained by using an eigenmode solver (Lumerical Solutions, MODE Solutions). To realize in- and outcoupling antennas, a shortcut between the two wire arms was created. In simulations this structure yields 30% incoupling efficiency for both modes if excited with a tightly focused Gaussian and NA = 1.4. This Further analysis of the mode-dispersions of the TWTL, the efficiencies, the mode conversion in the bending and the splitting can be found in the supplementary information.

**Sample Preparation:** Single-crystalline gold platelets were grown on a glass substrate by wet chemical synthesis[30]. A platelet with the desired thickness was covered with a PMMA droplet and moved onto a conductive substrate, where the PMMA was removed. Afterwards, the structures were fabricated by focused ion-beam milling with an Electron/Gallium-Ion dual-beam microscope (FEI Helios Dual-beam Nano-lab). To enhance optical characterization and to prevent the influence from residuals in the substrate from the ion-beam milling, the platelet with the structures was subsequently covered by a PMMA-droplet, and moved onto a flat pristine glass-substrate. Finally, the PMMA was removed and the structures were optical characterized through the glass substrate. For a sketch of the process, see the supplementary information. To accommodate small geometrical deviations during sample fabrication, a 2D array of structures was fabricated. Further details about the geometry of the structures can be found in the supplementary information.

**Optical measurements:** The optical measurements were performed with a home-built inverted microscope setup. A sketch of the setup is available in the supplementary information. A collimated 800 nm laser beam (λ = 800 nm, 12 nm FWHM spectral line width, 80 MHz repetition rate, NKT Photonics, SuperK Power with SpectraK AOTF, Master Seed pulse duration 5 ps, after AOTF about 300 ps), is linearly polarized (Lin. Polarizer B. Halle and λ/2 plate Foctech AWP210H NIR) and, after a beam-splitter (Thorlabs, CM1-BS2 – 30 mm, Non-Polarizing Beam-splitter, 50:50, 700 nm – 1100 nm), circularly polarized (λ/4 plate Thorlabs, WPQ10ME-808). The beam is tightly focused (Zeiss Objective Plan-Apochromat 63x / 1.40 Oil DIC) onto the sample, which is mounted onto a piezo stage (PI, P-572). The emitted signal is collected by the same objective and, after passing the λ/4 plate, reflected at the beam-splitter. In an intermediate image-plane, a chrome disc (OD2) on a glass substrate weakens the back-reflection from the excited antenna to prevent saturation of the following CCD chip. A linear polarizer (B. Halle) acts as an analyser, and the signal is imaged onto a CCD camera (Andor, DV887AC-FI EMCCD, cooled to -40° C).

**Acknowledgements**

The authors thank Paolo Biaogioni, Thorsten Feichtner, Peter Geisler, Johannes Kern, René Kullock and Björn Trauzettel for fruitful discussion. The authors gratefully acknowledge funding of the DFG via SPP1391 and HE5618/6-1.

**Author contributions**

B.H., E.K., G.R. and S.G. conceived the project. E.K. fabricated and transferred the structures. G.R. performed the FDTD simulations. E.K. and D.K. performed the optical measurements and analysed the data. All authors participated in discussing the data and writing the manuscript.

# Supplementary Information for the manuscript entitled

# "A Spin-Optical Nano Device"

## Content

# 1. Device Efficiencies

The simulated ratio between the intensity emitted by one output terminal into the collection objective $I_{Out}$ and the intensity of the Gaussian excitation beam $I_{In}$ can be directly extracted from the exciting and emitted beam in simulations and amounts to $I_{Out}/I_{In} = 1.5\%$.

To discriminate between losses occurring inside the device and during conversion between far- and near-fields, we separately consider different contributions. In addition to in- and out-coupling antenna efficiencies and the transmittance at the Y-junction, propagation losses as well as bending losses will occur. All following efficiencies were obtained by simulations (Lumerical Solutions, FDTD Solutions and Lumerical Solutions, MODE Solutions) of the geometry sketched in Figure S1a.

**In- and out-coupling antennas:** The shortcuts at the input (see Fig. S1b) and the output terminals act as efficient antennas even superior to dipole antennas[23]. The shortcuts are easy to fabricate, since they exhibit only few geometrical details. For the present geometry, we obtain an in-coupling efficiency of $E_{Ant,s} = 33\%$ for the symmetric as well as the antisymmetric mode when excited with a focused Gaussian beam (NA = 1.4). Linear dipole antennas as used in[15] only reach efficiencies on the order of 20%. In-coupling efficiency here means the ratio of power in the transmission line eigenmode in a short distance behind the shortcut to the power in the incoming Gaussian beam. Due to reciprocity we assume the out-coupling efficiency to be similar to the in-coupling efficiency.

**Y-Junction:** Discontinuities along a plasmonic waveguide cause back reflection of a certain fraction of the intensity of a propagating plasmon mode. This back reflection results in standing waves. Since the standing wave pattern depends on the effective wavelengths of the respective modes, the ratio of both modes, visualized by the blue trajectory on the Poincaré sphere, exhibits small oscillations (Fig. 1c of the main manuscript). For our geometry, the value of the reflectivity (transmission) at the Y-junction (see Fig. S1c) is 2%(119%) for the symmetric and 3%(63%) for the antisymmetric mode. Note that the transmission of the symmetric mode exceeds 100%, since mode conversion takes place.

**Propagation and bending losses:** Since mode-conversion takes place around the middle of the device, we estimate the propagation losses of each mode ($l_{decay,s} = 3.70$ μm, $l_{decay,as} = 2.28$ μm) for half of the total propagated distance ($l_{1/2 total} = 1.6$ μm) neglecting the bending. This yields a transmission of the modes intensity of $E_{Prop,s} = 65\%$ and $E_{Prop,as} = 50\%$. The transmission of the bending, including propagation losses along the bending, is $E_{Bending,s} = 32\%$ and $E_{Bending,as} = 108\%$.

**Overall efficiency:** To obtain the overall ratio of the power in the Gaussian excitation beam and the emitted power, we multiply all these factors. To obtain an estimate of the overall efficiency we take the mean value of the efficiencies of both modes (see Fig. S1d), since we assume equal intensities along the propagation. With $E_{Ant,in,mean} = E_{PAnt,out,mean} = 33\%$, $E_{Y,mean} = 91\%$ and $E_{Bending,mean} = 70\%$ we obtain

$$\frac{I_{Out}}{I_{In}} = E_{Ant,in,mean} * E_{Prop,s} * E_{Y,mean} * E_{Bending,mean} * E_{Prop,as} * E_{Ant,out,mean} = 2.2\%.$$

Note that this value differs from the initially stated value, since we assumed a completely homogeneous mode-distribution throughout the whole device, and neglect smaller contributions like higher-order reflections and interferences. While the value obtained directly from the power of the exciting and emitted beam is more accurate, by obtaining the individual contributions the overall efficiency can be divided into the external efficiency, i.e. the combination of the antennas in- and out-coupling efficiencies, and the internal efficiency, i.e. the remaining contributions. The antennas contributions alone lead to a transmission of only 11% of the intensity. Neglecting this far-field conversions, the internal efficiency of the device (including around $4\ \mu m$ of near-field propagation, mode conversion and discontinuities) is as high as 21%, highlighting its applicability in integrated circuits or in combination with, e.g. quantum emitters, where much more efficient coupling is possible.

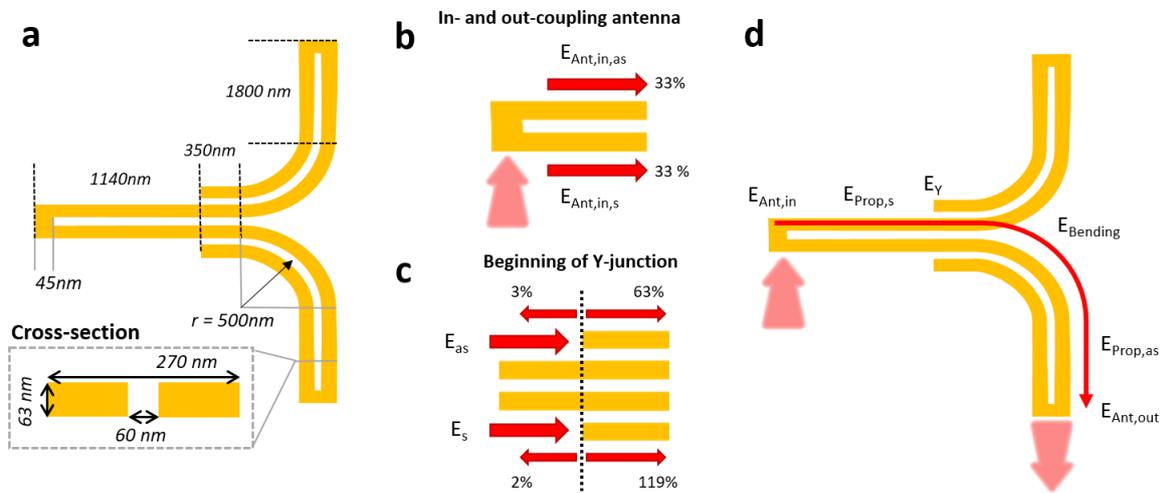

**Figure S1 | Waveguide efficiencies.** a) Geometry of the nano-device. b) In-coupling efficiencies for both modes for the used split-ring antenna design. c) Reflection and transmission of both modes at the Y-junction. Note that due to mode-conversion, the symmetric mode's transmission is above 100%. d) Sketch of the transmission efficiencies throughout the whole device.

## 2. Tuning of the Waveguide's Birefringence

The dimensions of the TWTL are chosen such that only two modes propagate significant distances along the TWTL (see Fig. S2a). Due to their differing spatial distribution and the resulting effective refractive indices, the modes differ in their damping and effective wavelength. For an operating wavelength of $\lambda_{vac} = 800\ nm$, this results in $\lambda_{eff,s} = 480.3\ nm$ and $l_{decay,s} = 3.70\ \mu m$ for the symmetric and $\lambda_{eff,as} = 438.8\ nm$ and $l_{decay,as} = 2.28\ \mu m$ for the antisymmetric mode. Equivalent to an optical birefringent medium, during propagation the two modes acquire a phase shift induced by the different effective wavelengths. For a linear infinitely long TWTL the superposition state describes a trajectory in the Poincaré representation that spirals towards the pure mode, which exhibits the smaller damping (in general the symmetric, less confined mode) as visualized in Figure S2b.

Expectedly, this trajectory is sensitive to changes that mainly influence the propagation of one mode. If, for example, the gap of the TWTL is filled with glass, the new mode properties are $\lambda_{eff,s} = 476.5\ nm$ and $l_{decay,s} = 3.47\ \mu m$ for the symmetric and $\lambda_{eff,as} = 369.7\ nm$ and $l_{decay,as} = 1.75\ \mu m$ for the antisymmetric mode, i.e. mainly the antisymmetric mode's dispersion is modified. For the same propagation length and identical circularly polarized excitation the trajectory on the Poincaré sphere (see Fig. S2c) strongly differs from the initial case. The evolution of the polarization state of the mode superposition can therefore be engineered by the dielectric environment. This may enable controlled switching if an optical active material is combined with the device.

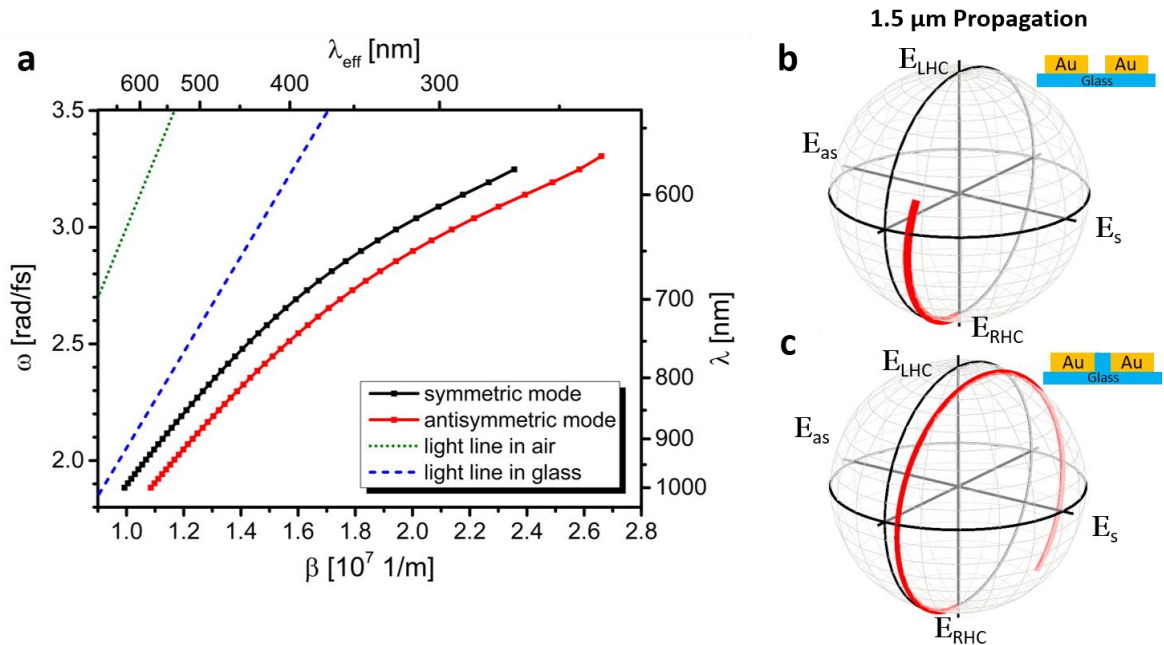

**Figure S3 | Waveguide birefringence.** a) Simulated dispersion relation of the TWTL eigenmodes for the cross section sketched in Fig. S1a. b,c) Trajectories of the polarization state on the Poincaré sphere of an initially circularly polarized mode superposition (trajectory starts at the south pole) for the cross-section sketched in Fig. S1a for a propagation length of 1.5 μm (b) as well as for the case with glass filling the TWTL gap (c).

## 3. Poincaré sphere representation

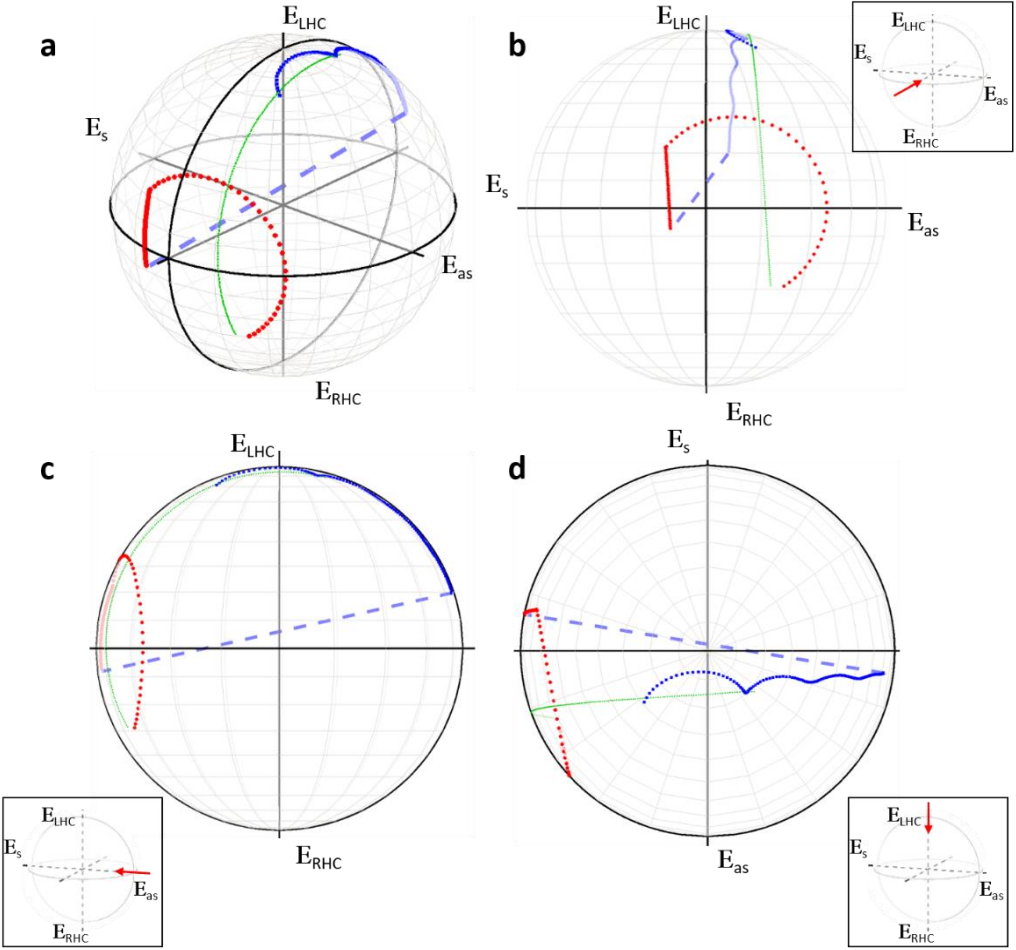

**Figure S2 | Poincaré sphere representation.** Visualization of the trajectory of the pseudo-spin on the Poincaré sphere throughout the whole nano-device (a), also shown from the front (b), side (c) and top (d) view. The different sections are indicated by the same colours as in Fig. 1b,c of the main manuscript.

# 4. Spatial Distribution of Mode Superposition

A superposition of the symmetric and the antisymmetric mode causes a mode beating along the transmission line that manifests itself in a periodic localization of the overall near-field intensity on either side of the TWTL. This can be utilized to direct the trajectory of plasmons at a junction[1,24]. The relative starting phase between both modes determines the resulting intensity distribution along the TWTL (see Fig. S4a for an excitation with left-handed circular polarization). This superposition also extends if two additional wires are added (see Fig. S4b). In principle, for the four-wire transmission line new fundamental modes need to be considered. However, as is seen by comparison of Figure S4a and Figure S4b, for short propagation distances the two outer wires have only marginal influence on the near-field intensity distribution. It is therefore possible to bend two wires on each side away, forming two new separated TWTLs. The wire pair on that side, where the main intensity is concentrated (here: the right side), then carries the main portion of the propagating plasmon (see Fig. S4c).

While the geometry of the device is optimized for an operating wavelength of 800 nm the bandwidth over which the splitting works is quiet broad. This is due to the fact that the position at which the splitting occurs is determined by the beating wavelength, which is quite large. The device therefore supports routing of ultrashort laser pulses[24].

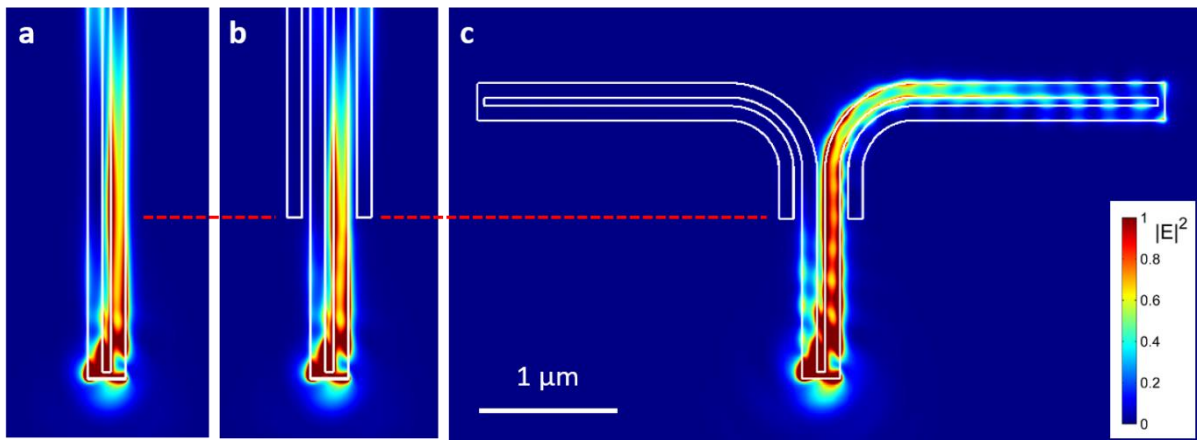

**Figure S4 | Mode Superposition.** a) The beating of the two fundamental modes of a TWTL results in a near-field intensity pattern along the transmission line for which the intensity oscillates from left to right. b) This pattern is unchanged for about 500 nm if two wire are added – one on each side. c) When the two pairs of wires are bend apart two new separated TWTLs are formed. For the chosen configuration the excitation mainly propagates into one of the two transmission lines. Simulated near-field intensity distributions are recorded in glass *10 nm* below the glass-air interface.

# 5. Mode-Conversion in the Bending

If the translation symmetry of the TWTL is broken the fundamental modes are no longer eigenmodes of the TWTL. Therefore, a conversion between modes is expected. Such mode conversion between the symmetric and the antisymmetric mode in a TWTL has already been reported[25]. Here, mode conversion is achieved by bending the TWTL. We studied such mode conversion for a variety of TWTL bends. Figure S5 demonstrates mode conversion at the example of a fixed bending radius as a function of the bending angle. For a bending angle of 90°, simulations of an excitation with only the antisymmetric (Fig. S5a) or symmetric (Fig. S5b) mode result in propagation of mainly the other mode after the bending. During propagation along the bend – parametrized by the bending angle - the intensity in the original mode (solid black and red line in Fig. S5c) nearly vanishes, while the intensity in the other mode increases (dotted black and red line in Fig. S5c). We note that the obtained transmission efficiencies depend on the superposition of modes that enters the bend.

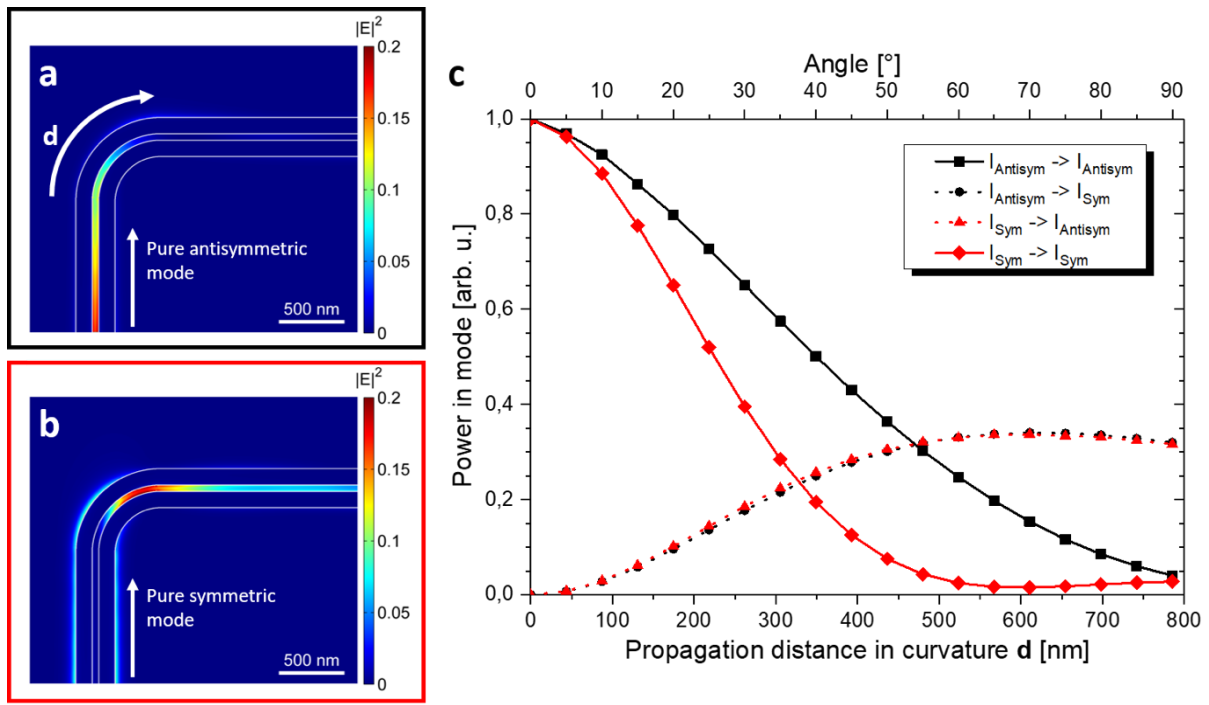

**Figure S5 | Mode conversion.** a,b) Simulated near-field intensity distribution along a TWTL bend for excitation by the pure antisymmetric (a) or symmetric (b) mode. c) Simulated intensity along the bend for a pure antisymmetric (black lines) and symmetric (red lines) mode input. The solid lines represent the intensity in the initial mode, while the dotted lines represent the intensity in the converted mode. Mode conversion is clearly observed.

# 6. Sample Preparation

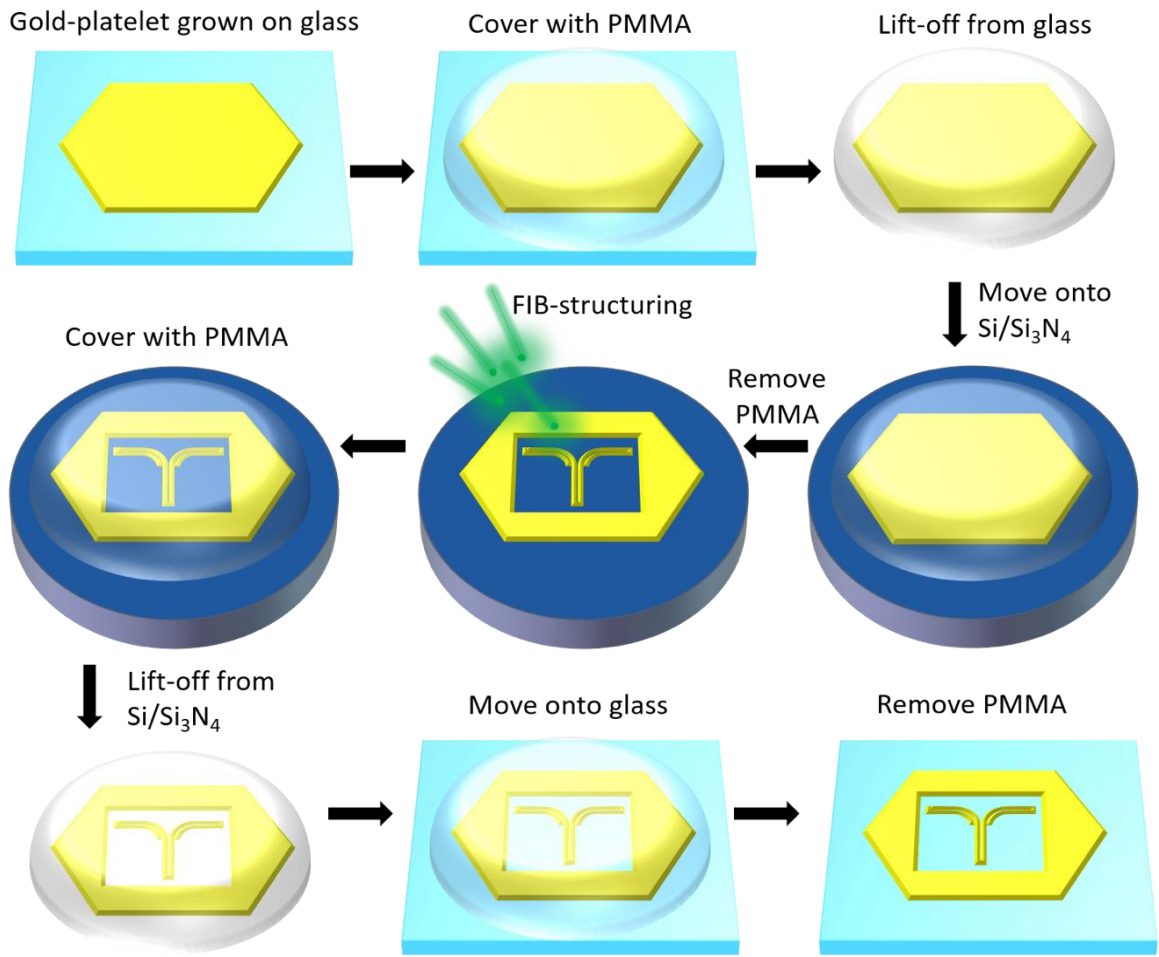

**Figure S6 | Sample preparation.** A single-crystalline gold-platelet is grown on a glass substrate. It is covered with a PMMA-droplet, removed from the glass substrate and moved onto a Si-Waver piece with a $Si_3N_4$ layer on top. The PMMA is removed and the structure is milled out of the gold-platelet by focused ion-beam milling. The structured gold platelet is then covered again by a PMMA-droplet, removed from the Si-Waver piece and moved onto a pristine glass substrate. Finally before experiments the PMMA is removed.

# 7. Optimization of the Waveguides Geometry

An initial efficient geometry of the device was obtained by optimization in FDTD simulations. The final geometry of the realized structure differed from this optimized geometry due to multiple reasons. In simulations a rectangular wire cross-section is assumed. In reality, the cross-section of the experimentally investigated structure is determined by the beam-profile of the focused ion-beam and therefore will deviate from a rectangular profile. The used gold platelet is *63 nm* thick, which we assumed to be the final structures height. High-resolution SEM images of the final structures obtained on the glass substrate indicate an antenna shortcut width of *45 nm*, a gap width of about *60 nm* and a total TWTL width of *270 nm* (see also Fig. S7a). To compensate the influence of these geometrical deviation from the initial simulations, the propagation distances before and after the Y-Junction and the bending ($L_{inital}$ and $L_{final}$ in Fig. S7a) were changed, while all other parameters were kept fixed. This way the phase delay between both modes can be tuned. An array of structures was fabricated with $L_{inital}$ being varied from *740 nm* to *1340 nm* and $L_{final}$ from *1200 nm* to *1800 nm* both in steps of *200 nm*. The fabricated structure that, in experiments, came closest to the desired results had an initial length $L_{inital}$ of *1140 nm* and a final length $L_{final}$ of *1800 nm*. These dimensions were then used for the simulations shown in the main manuscript as well as the supporting information. The fact that we were able to compensate the deviations in TWTL cross-section and antenna efficiencies by changes in the modes' propagation distance highlights the robustness of our design.

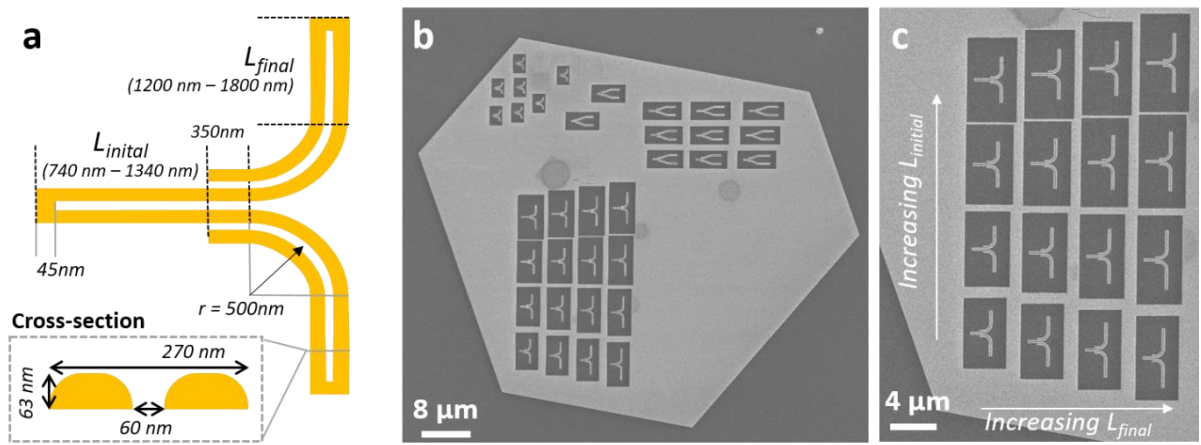

**Figure S7 | Sample Layout.** a) Geometrical parameters for the realized structures. One single-crystalline platelet (b) with a uniform thickness of 63 nm was used to structure the array of the nano-devices (c). In this array, the lengths of both initial and final linear propagation segments were scanned to compensate small deviations during structuring. Other geometrical parameters (e.g. radius of the curvature, TWTL cross-section) were kept fixed.

# 8. Optical Setup and Measurements

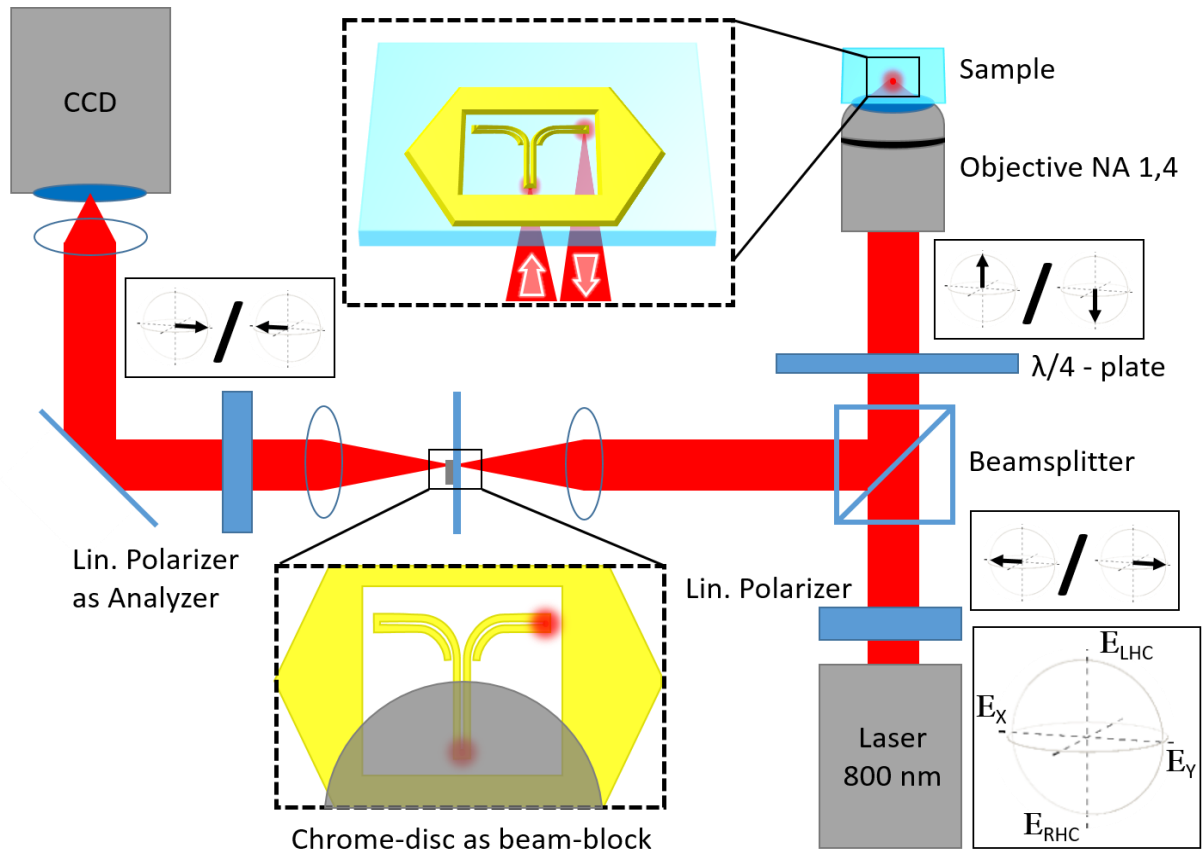

**Figure S8 | Optical Setup.** A linearly polarized 800nm laser beam passes a λ/4 plate, resulting in a circular polarization. Afterwards, the beam is tightly focused (NA 1.4) onto the sample, which is positioned in the focus with a piezo stage. The emitted signal as well as the reflection pass the λ/4 plate and are reflected at a non-polarizing beam-splitter. In an intermediate image-plane, a chrome disc attenuates the directly reflected light from the excitation spot. A linear polarizer acts as an analyzer, and the signal is imaged onto a CCD-camera.